\documentclass[ reprint,showpacs,
 superscriptaddress,
 pra,
]{revtex4-1}
\usepackage{amsfonts}
\usepackage{mathrsfs,amssymb}
\usepackage[top=1in, bottom=1in, left=1in, right=1in]{geometry}
\usepackage{amsmath}
\usepackage{epsfig}
\usepackage{bm}
\usepackage{graphicx}
\usepackage{grffile}
\usepackage{ulem}
\usepackage{cancel}
\usepackage{cases}
\usepackage{mhchem}
\usepackage{mathtools}

\renewcommand{\vec}[1]{\bm{\mathrm{#1}}}
\begin{document}

\title{Quantum storage based on controllable frequency comb}
\author{Xiwen Zhang}
\email{xiwen@physics.tamu.edu}
\affiliation{Texas A\&M University, College Station, Texas 77843, USA}
\author{Alexey Kalachev}
\affiliation{Zavoisky Physical-Technical Institute of the Russian Academy of Sciences, Sibirsky Trakt 10/7, Kazan, 420029, Russia}
\author{Philip Hemmer}
\affiliation{Texas A\&M University, College Station, Texas 77843, USA}
\author{Olga Kocharovskaya}
\affiliation{Texas A\&M University, College Station, Texas 77843, USA}


\date{\today }

\begin{abstract}
We suggest an all-optical scheme for the storage, retrieval and processing of a single-photon wave packet through its off-resonant Raman interaction with a series of coherent control beams. These control beams, each with distinct carrier frequency, are distributed along the way of single-photon propagation, thus effectively forming a gradient absorption structure which can be controlled in various ways to achieve different single-photon processing functionalities. Such a controllable frequency comb is a hybrid of Raman, gradient echo memory (GEM) and atomic frequency comb (AFC) methods, therefore demonstrates many of their advantages all together in one.
\end{abstract}


\maketitle
After three decades of endeavor, quantum information processing has been developed from visionaries' conceptual idea~\cite{Feynman82} into a vast diversity of research subjects, among which quantum memory draws lots of attention and has undergone a huge development in the last ten years because of the important role it plays in quantum computer and long-distance quantum communication~\cite{Lvovsky09, Sangouard11}, etc. The major techniques for the storage and retrieval of a time-bin single-photon wave packet~\cite{Brendel99} in atomic ensembles divide into two groups. The first one applies an optimized shaped-in-time control field to convert a single photon into collective spin waves. This includes electromagnetically induced transparency (EIT)~\cite{Eisaman05, Chaneliere05} and off-resonant Raman scheme~\cite{Reim11, Bustard13, England13}, which requires prior knowledge of the single-photon waveform for optimization and its arrival time for signal-control syncretization. The second category takes advantage of photon-echo mechanism, such as revival of silenced echo~\cite{Damon11}, controlled reversible inhomogeneous broadening~\cite{Moiseev01} and/or gradient echo memory (GEM)~\cite{Hetet08, Hosseini09, Sparkes13}, atomic frequency comb (AFC)~\cite{deRiedmatten08, Afzelius09, Afzelius10, Saglamyurek14}, etc. They usually require a use of $\pi$ pulses, the existence of Stark and/or Zeeman effect, or delicate spectral tailoring on a broad inhomogeneous broadening. Nevertheless, these requirements of the quantum memory schemes can become limiting factors for the actuarial implementation in different light-matter interfaces under various scenarios. Here we propose a method combining these two types of techniques, which overcomes the above hassles and demonstrates many other interesting advantages.

\begin{figure}[h]
\begin{center}
\epsfig{figure=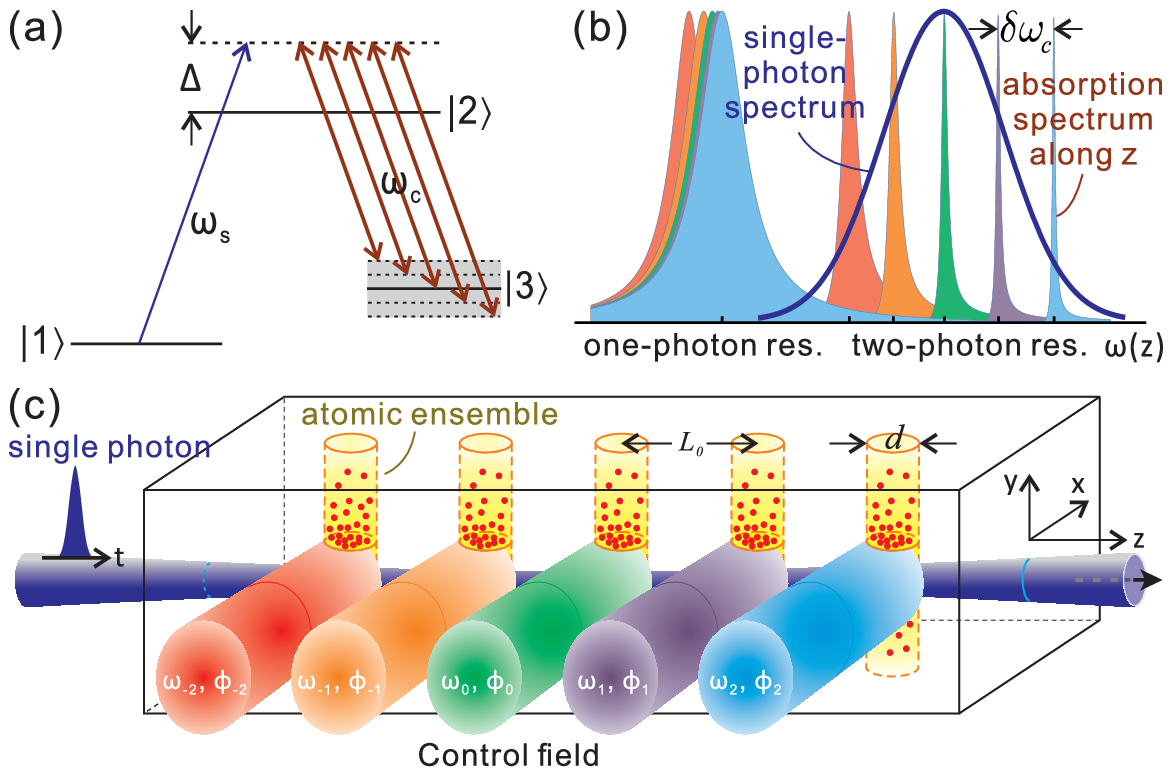, width=7.8cm}
\end{center}
\caption{Scheme of quantum memory based on controllable frequency comb. (a) In a $\Lambda$-level structure, the control field consists of a number of spatially separated beams each with a different carrier frequency. (b) Illustration of the imaginary part of the susceptibility in different sections of the medium, where the two-photon resonance contribution effectively forms a frequency comb distributed along $z$ direction. (c) Possible implantation of the controllable frequency comb.}
\label{FigureScheme}
\end{figure}

Consider a single-photon wave packet of carrier frequency $\omega_s$ and duration $\Delta t$ passing through a three-level atomic ensemble along the longitudinal direction $\hat{z}$ with one-photon detuning $\Delta$ [Fig. \ref{FigureScheme}(a)].
The control field propagates perpendicularly to the longitudinal direction, with its frequency distributed across the field, which is called spatial chirp. The interaction volume consists of $M$ discrete sections (each with length $d$ and inter distance $L_0$) due to spatial discontinuity of either atomic ensemble or control beams [Fig. \ref{FigureScheme}(c)]. Because of the control field spatial chirp, the single photon meets different two-photon resonance conditions at different sections of the medium [Fig. \ref{FigureScheme}(b)], resulting in a frequency comb in its absorption spectrum. But unlike AFC, the comb teeth are distributed linearly along the way of photon propagation, therefore forming a gradient frequency comb (GFC).

The signal and control fields are written as~\cite{Milonni95}
\begin{align}
\vec{E}_{s}\left( z,t\right) & =\vec{\epsilon}_{s}\frac{i}{n_{\text{bg}}}\mathscr{E}_{s}a\left( z,t\right) e^{i(k_{s}z-\omega _{s}t)}+H.c., \\
\vec{E}_{c}\left( \vec{r},t\right) & =\vec{\epsilon}_{c}E_{0}(z)e^{i[k_{c}(z)x-\omega _{c}(z)t-\varphi _{0c}(z)]}+c.c.,
\end{align}
respectively, in which $n_\text{bg}$ is the background refractive index, $a\left( z,t\right) $ is the slowly varying amplitude of the signal field, $\mathscr{E}_{s}=\sqrt{\frac{\hbar \omega _{s}}{2\varepsilon_{0}V}}$, $k_{c}=n_{\text{bg}}\omega _{c}/c$, $k_{s}=n\omega_{s}/c$, $n$ is the refractive index taking into account both background and Raman interaction, $c$ is the speed of light in vacuum, $\varphi _{0c}$ is a stationary phase on the control beam. We assume each control beam has much longer duration compared with the signal, so that its amplitude $E_0(z)$ is treated to be time invariant. The control field spatial chirp is characterized by its position-dependent frequency: $\omega_{c}(z)-\omega_{c0}=m\delta \omega_{c}$, where $\omega_{c0}=\omega_{c}(z=0)$, $m=m(z)$ enumerates the number of control beams, and $\delta \omega_{c}$ is their adjacent frequency spacing.

Let us define the slowly varying part of the spin wave operator as
\begin{equation}
S\left( \vec{r},t\right) \propto g^* N \sigma_{13} e^{i\left( \omega _{s}-\omega_{c}\right) t}e^{i\left[ \vec{k}_{c}(z)-\vec{k}_{s}\right] \cdot \vec{r}}e^{-i\varphi _{0c}} , \label{S}
\end{equation}
where $\sigma _{nn^{\prime }}(\vec{r},t)=\frac{1}{N}\sum_{j}\left\vert n\right\rangle_{j}\left\langle n^{\prime }\right\vert \delta (\vec{r}-\vec{r}^{j})$ is the collective atomic operator, $\left\vert n\right\rangle_j$ is the $n^\text{th}$ state of atom $j$, $N$ is the atomic density, $g=\sqrt{\frac{\omega_{s}}{2\hbar \varepsilon_{0}c n}}\frac{d_{21}\Omega_{c}(z) }{\Delta }$ is the coupling constant, $d_{21}$ is the corresponding transition matrix element, $\Omega_{c}$ is the control field Rabi frequency. In the long-pulse and far-off-resonant regime, the evolution equations are given by
\begin{align}
\frac{\partial }{\partial z}a\left( z,t\right) & =-S(z,t) ,  \label{Eq. a} \\
\frac{\partial }{\partial t}S\left( z,t\right) & =-\left[ \gamma_{31}-i\delta (z)\right] S\left( z,t\right) +|g|^{2}N a\left(z,t\right) ,  \label{Eq. S}
\end{align}
where $\gamma_{31}$ is the spin wave decoherence rate. We choose the control field central frequency as $\omega _{c0}=\omega _{s}-\omega _{31} -|\Omega_{c}(z=0)|^{2} / \Delta$ ($\omega_{31}$ is the spin transition frequency), then the two-photon detuning becomes
\begin{equation}
\delta (z)=-\sum_{m=-M_{0}}^{M_{0}}m\delta \omega _{c}(\Theta_{-}^{m}-\Theta _{+}^{m})-\delta _{AC}, \label{delta}
\end{equation}
where $\Theta_{\mp }^{m}=\Theta( z-mL_{0}\pm d/2 )$ are the heaviside step functions, and $\delta_{AC}(z)=[ |\Omega_{c}(z)|^{2} -|\Omega_{c}(z=0)|^{2}] / \Delta$ is the uncompensated ac-Stark shift due to the spatial inhomogeneity of each control beam. Depending on these control beam's intensities and waists, the variation of $\delta_{AC}$ within each section of the medium can become on the same order of, or even larger than, the input photon bandwidth, in which case the retrieved signal will be strongly degraded. Therefore in the following we always assume control beams with either flat-topped profile or large enough waist, such that $\delta_{AC}\approx0$.

\begin{figure}[h]
\begin{center}
\epsfig{figure=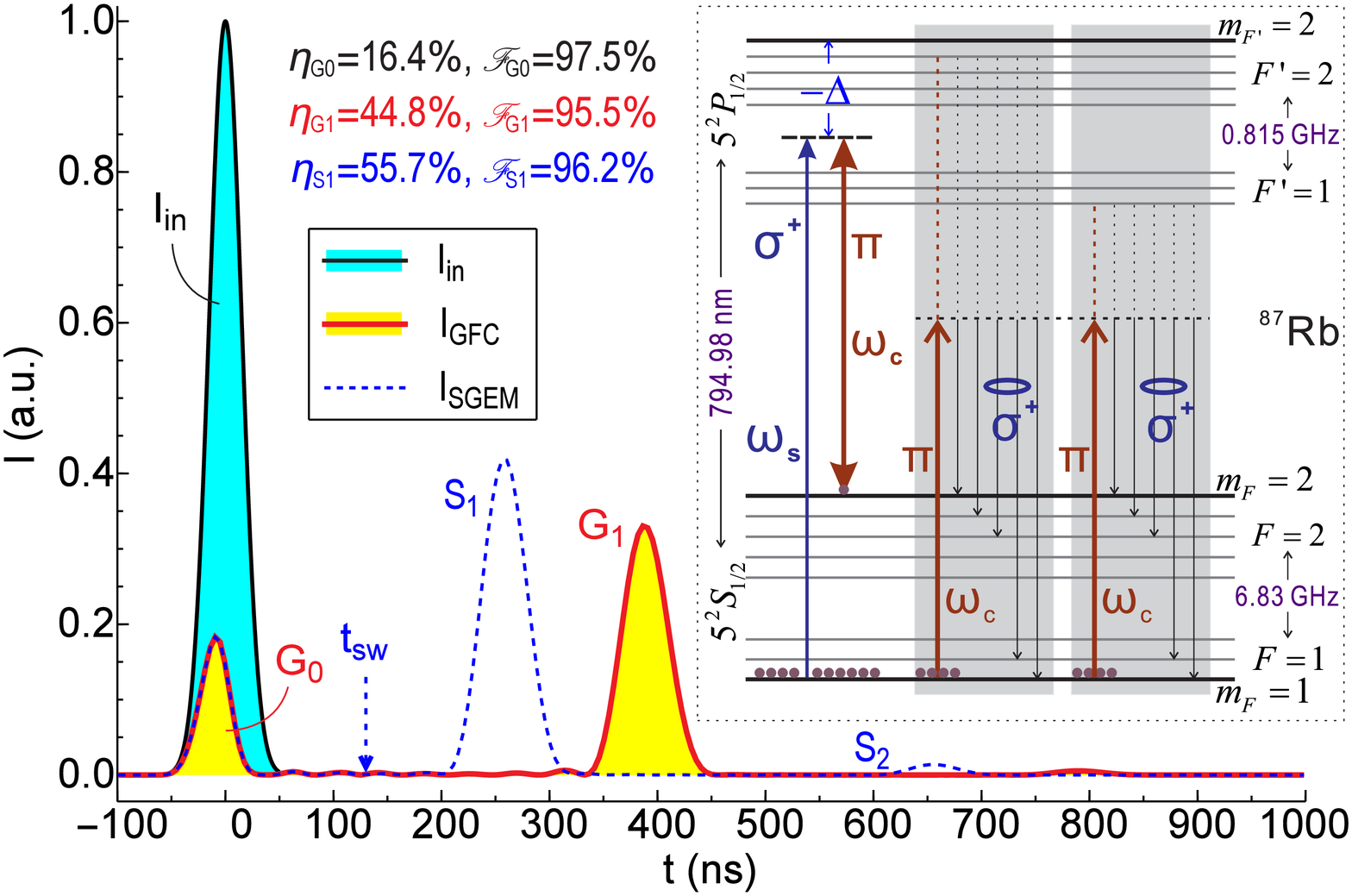, width=7.9cm}
\end{center}
\caption{GFC (red line, yellow filled) and SGEM (blue dashed line) echoes in a cold $^{87}$Rb ensemble of density $10^{11}$~cm$^{-3}$ and temperature $100$~$\mu$K. The inset shows the involved energy levels, with $\Lambda$ scheme consisting of $\left\vert F=1,m_F=1\right\rangle$-$\left\vert F'=2,m_{F'}=2\right\rangle$-$\left\vert F=2,m_F=2\right\rangle$ of rubidium D1 line. The medium is divided into $M=9$ sections each of length $d=0.56$~mm. The $\sigma^+$-polarized signal photon ($\Delta t = 50$~ns, width $50$~$\mu$m) is frequency detuned by $-0.7$~GHz from the excited state. The nine $\pi$-polarized control beams of total power $0.18$~W are assigned to a frequency spacing such that $T_0$ = 400~ns, giving $\zeta^0_\text{eff} = 1.28$. In SGEM regime, $T_\text{sw}=130$~ns. The shaded area in the inset shows the major noise channels.}\label{FigureColdAtom50ns}
\end{figure}

The Raman interaction excites spin waves in the atomic medium during storage process. Due to the periodicity of the two-photon resonance condition, the spin waves recurrently get in-phase after each $T_0=2\pi / \delta \omega_c$. Since the coherence for spin transitions are usually long-lived, the comb finesse $\mathcal{F}=\delta \omega_c / (2\gamma_{31}) \gg 1$ is assumed to be always satisfied. In such a case Eqs. (\ref{Eq. a}) and (\ref{Eq. S}) are solved analytically up to $t=T_0$ as follows~\cite{Zhang15TBP}:
\begin{align}
a_{\text{out}}(t)& \approx e^{-\frac{\pi }{4}\zeta_{\text{eff}}^{0}}a_{\text{in}}(t)-\frac{\pi \zeta_{\text{eff}}^{0}}{2}e^{-\frac{\pi \zeta_{\text{eff}}^{0}}{4}}e^{-\frac{\pi }{\mathcal{F}}}a_{\text{in}}(t-T_{0}),
\label{SlnI}
\end{align}
where $\zeta^0_{\text{eff}}=4|g|^2Nd/\delta \omega_c$ is the effective optical thickness of each section of the medium. We use efficiency $\eta$ to describe the ratio of retrieved energy out of the input, and fidelity $\mathscr{F}$ to characterize the waveform preservation~\cite{Zhang13}. From Eq. (\ref{SlnI}), the optimization condition can be obtained as $\zeta^0_{\text{eff}}=4/\pi$ and $\Delta t<T_0<M\Delta t$ for $\mathcal{F} \gg 1$. Under this condition, the forward GFC echo reaches the maximum efficiency $54\%$ (limited by the reabsorption processes), demonstrating the same action as AFC scheme.

As an example, let us consider the $\Lambda$-level structure depicted in Fig. \ref{FigureColdAtom50ns} in cold $^{87}$Rb ensemble with atomic density $10^{11}$~cm$^{-3}$~\cite{Sparkes13}. In such a case, a total control power of $0.18$~W yields a $|g|^2N = 9\times 10^9$~s$^{-1}$m$^{-1}$ under a one-photon detuning $\Delta/(2\pi) = -0.7$~GHz, which is enough for storing a signal field of $\Delta t = 50$~ns. The control beams can be generated, for instance, by passing a cw laser field through an acousto-optic modulator and arbitrary wave generator, which permit a bandwidth of $\sim 100$~MHz. The detuning is chosen on the one hand to maintain reasonably low noise level~\cite{Zhang14PRA}, on the other hand to satisfy off-resonant Raman condition~\cite{Gorshkov07(2)}. There are ten major noise channels in this case, which add up to give an operation window $<1.21$~$\mu$s in order to keep signal-to-noise ratio smaller than $1$ without filtering. Since six of them produce noises with polarizations different from the signal, a simple polarization filter will enlarge this time window to $3$~$\mu$s.
The spin wave Doppler dephasing along the longitudinal frequency gradient due to atomic motion with Maxwellian velocity distribution~\cite{Zhao08} is taken into account by multiplying a Gaussian factor $\exp [ -t^{2}/(2t_{d}^{2})]$ onto the right-hand-side of Eq. (\ref{Eq. a}), where $t_{d}=1/(\sqrt{k_{B}T/m_a}k_{s})$ is the dephasing time, $k_{B}$ is the Boltzmann constant, $T$ is the gas temperature and $m_a$ is the atomic mass. The typical temperature of such system is $100$~$\mu$K, which gives a dephasing time $t_d=1$~$\mu$s and atom-loss time $\sim 78$~$\mu$s, longer than the storage time $T_0=400$~ns. In general, a transverse atomic motion also leads to similar dephasing of the spin wave, which will be neglected for the sake of simplicity.

\begin{figure}[h]
\begin{center}
\epsfig{figure=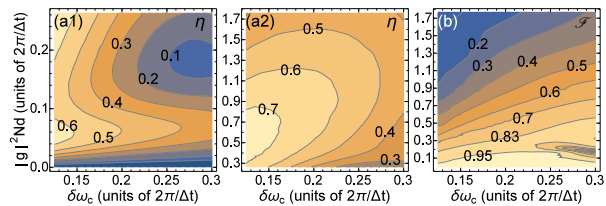, width=7.95cm}
\end{center}
\caption{Numerical simulation of SGEM first echo efficiency (a1, a2) and fidelity (b) based on Eqs. (\ref{Eq. a})-(\ref{delta}). Parameters are $M=9$, $\Delta t = 50$~ns, $\gamma_{31}=0$~rad/s, and $T_\text{sw} = 130$~ns.}\label{FigureSGEMeffContour}
\end{figure}

\begin{figure}[h]
\begin{center}
\epsfig{figure=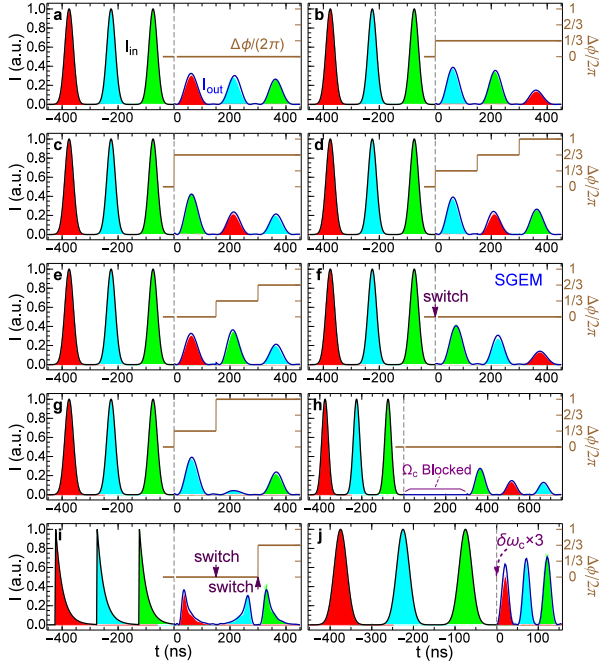, width=7.9cm}
\end{center}
\caption{Single-photon processing. The black and blue solid lines represent input ($t<0$~ns) and output ($t>0$~ns) triple-peak single photon intensities, respectively. The filled colors corresponds to single-peak signals. The brown solid lines show additional phased difference $\Delta \phi / (2\pi)$ modulation. (a-f) Sequencing signals with (a-c) circular and (d-f) non-circular permutations, where in (f) instead of using phase modulation we take advantage of SGEM regime. (g) Retrieving a triple-peak photon without the first peak. (h) Storing a photon for additional $300$~ns by blocking the control beams. Since the control field has a phase evolution during this time, the recalled signal switches the sequence accordingly. (i) Retrieving a photon with the second peak reversed while preserving the original temporal sequence. When adding additional phase (brown solid line), it is worth noting that here the frequency switching (indicated by the short arrows) is assumed to be achieved by swapping the control beams of opposite carrier frequencies, which inherently causes a phase shift if $t_\text{sw}\neq nT_0$. (j) Compressing a photon via increasing the control beam frequency spacing by three times. Common parameters: $\Delta t = 50$~ns, $T_0 = 450$~ns, $M=9$, $\zeta^0_\text{eff} = 4/\pi$, except that in (i), $M=31$ to catch the sharp-edge feature of the input photon, and in (j), $T_0 = 450\text{~ns} \to 150\text{~ns}$, $\zeta_\text{eff}^0 = 6/\pi \to 2/\pi$. Parameters for material are the same as in Fig. \ref{FigureColdAtom50ns}. }\label{FigurePulseSeq}
\end{figure}

The fact that our frequency comb is implemented through two-photon resonance by external control beams gives us additional degrees of freedom for the manipulation of a single-photon wave packet. First, a switch of the control beams' frequencies to the opposite (with respect to $\omega_{c0}$), at a moment $t_\text{sw}=t_\text{in}+T_\text{sw}$, effectively reverses the stepwise frequency gradient of the comb, resembling GEM scheme in a discrete manner, called stepwise gradient echo memory (SGEM) (see Fig. \ref{FigureColdAtom50ns}). In this regime, the first echo is formed due to the rephasing, rather than beating, of the spin waves in the medium, with controllable storage time $2T_\text{sw}$. Since the resonance condition is longitudinally distributed over space, the rephasing process is not strongly limited by the reabsorption of the echo. Therefore, it allows an on-demand retrieval of the input photon in a time-reversed order [Fig. \ref{FigurePulseSeq} (f, i)] with efficiency higher than the theoretical limit of the forward GFC echo ($54\%$), as shown in Fig. \ref{FigureSGEMeffContour}.

Second, as seen from Eq. (\ref{S}), an additional phase modulation of the control beams modifies the collective operator $S$, shifting and permuting the echo in time. Without this phase modulation, the emergence time of the echo is determined by the frequency spacing between nearby control beams. If extra phases are imposed to the control field in such a way that, between two adjacent beams the additionally introduced phase difference is $\Delta \phi = \delta \omega_c \tau$, $\tau \in [0,T_0)$, the phase evolution of the spin waves will be advanced by a time $\tau$, changing the first echo's appearance time to $t_\text{in}+T_0 - \tau$. This can be used for temporal sequencing of a single photon, as shown in Fig. \ref{FigurePulseSeq} (a-e).

Third, an increase (decrease) of the control beams' frequency spacing $\delta \omega_c$ during retrieval will accordingly reduce (magnify) the rephasing time and compress (stretch) the recalled signal [Fig. \ref{FigurePulseSeq} (j)], which is important for high bitrate transmission. In addition to the above mentioned, one can also achieve many other functionalities, for example: By combining phase modulation and frequency switching, we can flip the temporal shape of a specific time bin [Fig. \ref{FigurePulseSeq} (i)]. By offsetting the frequency of the write and/or read control field, one can modulate and shift the frequency of the retrieved signal, therefore realize a continuously tunable frequency conversion interface of a single photon for the purpose of multiplexing, routing, etc. All these manipulations can be delayed for arbitrary amount of time (allowed by the spin wave decoherence) via blocking, or turning off, the control beams [Fig. \ref{FigurePulseSeq} (h)]. Moreover, the techniques used in AFC, such as backward retrieval~\cite{Afzelius09, Zhang13} (for near-$100\%$ efficiency), time-to-frequency multiplexing, temporal and spectral filtering~\cite{Saglamyurek14}, etc, are as well applicable in our scheme.


The controllable frequency comb we propose is a hybrid of AFC and GEM in the framework of Raman configuration.
In AFC scheme, the comb bandwidth is determined by the inhomogeneous broadening and hyperfine splitting, and the finesse is limited by the imperfect optical pumping. The latter strongly reduces storage efficiency due to the background absorption, especially in optically dense medium.
The on-demand retrieval is done by transferring the atomic excitation down to another ground state using a $\pi$ pulse~\cite{Afzelius09, Afzelius10}, which can be problematic when the exact pulse area is difficult to prepare.
Since the absorption structure can not be much manipulated after being created, the signal processing has to be achieved by passing a frequency modulated single photon through a multiple AFC made in advance~\cite{Saglamyurek14}, which consumes the storage bandwidth, and may not be feasible in some situations since it makes a direct manipulation of the fragile single photon. In our case, the bandwidth and finesse are determined by the external controllability of the spatial chirp, the on-demand retrieval requires neither $\pi$ pulses nor a third ground state for spectrum tailoring, and pulse sequencing does not have the above problems.
In GEM scheme, the material has to demonstrate Zeeman or Stark effect for the creation of the frequency gradient, while our scheme is all-optical and can be implemented in a system possessing neither of these effects, or placed in an external fields for other purposes. In comparison with Raman scheme, we do not require prior knowledge of the single-photon waveform in order to have it stored.
So far AFC has been only implemented in rare-earth-doped crystals under cryogenic temperature. The above features of our scheme may allow the realization of AFC-like quantum memory in a variety of materials, such as atomic gases~\cite{Sparkes13} and color centers~\cite{Zhang14PRA}, or even room-temperature molecular ensembles~\cite{Bustard13, England13}.

We gratefully acknowledge the support of the National Science Foundation (Grant No. PHY-$130$-$73$-$46$). X.Z. is supported by the Herman F. Heep and Minnie Belle Heep Texas A$\&$M University Endowed Fund held/administered by the Texas A$\&$M Foundation. A.K. acknowledges the support from the Russian Science Foundation (Grant No. 14-12-00806).
\bibliographystyle{apsrev4-1}
\bibliography{QMDSCbibFile}

\end{document}